\documentclass[12pt,a4paper]{article}
\usepackage[vcentermath,enableskew]{youngtab}

\def\Dots{\cdot\cdot}
\def\QF{Q^N_F}

\begin{document}
\title{\hfill\raise 20pt\hbox{\small DIAS-STP-03-09}\\
Fuzzy Complex Quadrics and Spheres}
\author{Brian P. Dolan,$^{a),b)}$\footnote{\tt bdolan@thphys.may.ie}\hskip 10pt 
Denjoe O'Connor$^{b)}$\footnote{\tt denjoe@stp.dias.ie}\hskip 10pt and 
P. Pre\v{s}najder$^{c)}$\footnote{\tt presnajder@fmph.uniba.sk}\\
\\
$^{a)}${ Dept. of Mathematical Physics, NUI, Maynooth, Ireland}\\
\\
$^{b)}${ School of Theoretical Physics,}\\
{ Dublin Institute for Advanced Studies,}\\ 
{ 10~Burlington Rd., Dublin 8, Ireland}\\
\\
$^{c)}${ Dept. of Theoretical Physics, Comenius University,}\\ 
{ Mlynsk\' a dolina, SK-84248 Bratislava, Slovakia}}

\maketitle

\begin{abstract}
A matrix algebra is constructed which consists of the necessary degrees
of freedom for a finite approximation to 
the algebra of functions on the family of orthogonal Grassmannians
of real dimension $2N$, known as complex quadrics.
These matrix algebras contain the relevant degrees of freedom
for describing truncations of harmonic expansions of functions
on $N$-spheres.
An In\"on\"u-Wigner contraction of the quadric gives the
co-tangent bundle to the commutative sphere in the continuum limit.
It is shown how the degrees of freedom for the sphere can be projected out
of a finite dimensional functional integral,
using second-order Casimirs, giving a well-defined procedure for
construction functional integrals over fuzzy spheres of any dimension.

\end{abstract}

\section{Introduction}
Non-commutative geometry has slowly been increasing in importance
in physics over the last 20 years and has recently received
a strong impetus through work in string theory.  An important concept
in the non-commutative programme is that of a ``fuzzy'' space ---
this is perhaps more correctly described as a finite,
non-commuting, matrix approximation
to the algebra of functions on a continuous manifold which can reproduce 
the commutative algebra in the limit of the matrices becoming infinite in
size.  It has been suggested that fuzzy spaces could provide a
regularisation technique for numerical calculations in quantum field
theory which would be an alternative to lattice gauge theory,
\cite{GrosseKlimcikPresnajder}-\cite{Bal_etal}.
The prototypical example of a fuzzy space is the fuzzy two-sphere,
\cite{Madore}, but there are many more examples, and indeed
much of the work on generalised coherent states in quantum mechanics 
\cite{Perelomov}, when restricted to compact groups, 
can be related to fuzzy spaces.  
The geometry of fuzzy $CP^N$ and 
fuzzy unitary Grassmannians has been examined in the literature
in some detail \cite{Bordemann}-\cite{G2N}.  

Clearly, from the point of view of a regularisation technique in field theory
as well as for the theory of $D$-branes in string theory,
it would be desirable to have an explicit construction of the fuzzy
sphere $S^N_F$ in dimensions other than $N=2$.
Unfortunately, despite the elegant 
simplicity of the fuzzy two-sphere $S^2_F$, higher dimensional spheres are  
not so amenable to a fuzzy description.  To our knowledge there is no
closed finite matrix approximation to the algebra of functions on
a sphere for dimensions greater than two, though the problem was tackled
in \cite{Ramgoolam} and \cite{Kimura} (non-commutative spheres in the
continuum were analysed in \cite{ConnesDV}).  It
appears that, while one can represent truncated harmonic expansions
of functions on spheres by square matrices, the product of two such 
matrices takes one out of the required space and a projection
back into the space of functions is necessary after every
multiplication --- this renders the product non-associative.
In other words a star product cannot be defined on the fuzzy
sphere $S^N_F$ for $N>2$ (this is related to the fact that,
for $N\ne2$, $S^N$ cannot be obtained as the co-adjoint orbit of
a compact group).
Nevertheless the construction in \cite{Ramgoolam} does associate
a square matrix with the truncation of a harmonic expansion
on $S^N$ and so does, in a sense, constitute a fuzzy sphere,
even though there is no associative product.

For $S^4_F$ alternative descriptions are possible, \cite{GP} \cite{DJ}. 
The construction in \cite{DJ}
is in terms on $CP^3_F$ and uses the fact that the continuum
$CP^3$ is an $S^2$ bundle over $S^4$ \cite{DJ}.  
This approach has the
advantage that it is designed to be used in a functional integral and
the technique was extended to $S^3_F$ and $S^1_F$ in \cite{S3}.
The method of \cite{Ramgoolam} would be particularly cumbersome
to implement in functional integrals over $S^3_F$ and $S^1_F$,
or indeed any odd sphere.

In this paper we generalise the construction in \cite{S3} to
fuzzy spheres of any dimension.  The analysis relies
on properties of the orthogonal Grassmannian, $SO(N+2)/[SO(N)\times SO(2)]$,
which is a co-adjoint orbit of dimension $2N$.
This orthogonal Grassmannian
can also be obtained as the complex quadratic $z^az^a=0$ in $CP^{N+1}$,
where $z^a$ are natural complex co-ordinates, \cite{K+N}.
The notation in Kobayashi and Nomizu is 
$Q^N\cong SO(N+2)/[SO(N)\times SO(2)]$ and we shall
use this as a shorthand.
In\"on\"u-Wigner contraction
of \hbox{$SO(N+2)$} to the Euclidean group of ${\bf R}^{N+1}$ relates
$Q^N$ to the co-tangent bundle $T^*S^N$.
In a sense, elaborated on in section 2, $Q^N$
can be thought of as the compactified co-tangent bundle for
a non-commutative sphere.
There is a finite dimensional, non-commutative, matrix algebra
approximation to the algebra of functions $Q^N$
which contains harmonic expansions on $S^N$ (described in section 3).
While it is still the case that
the resulting matrix algebra takes one out of the space of
harmonic expansions on $S^N$ upon multiplication
(so one does not have a closed matrix approximation
to the algebra of functions on $S^N$ for $N\ne 2$)
it is nevertheless very easy to suppress unwanted modes in
a functional integral over a fuzzy complex quadric, $Q^N$,
in a manner which lends itself naturally to numerical
computation for field theory on $S^N_F$, as described
in section 4.

\section{The deformed co-tangent bundle}

In this section it is shown how the co-tangent bundle $T^*S^N$ 
can be deformed to a version related to a non-commutative sphere
and compactified to complex quadric,
$Q^N$. 

The construction starts with Cartesian co-ordinates $X^a$ in ${\bf R}^{N+1}$, where $a=1,\ldots,\hbox{$N+1$}$.
Consider the sphere $S^N\cong SO(N+1)/SO(N)$ of radius $R$ defined by
$X^aX^a=R^2$. 
The isometry group is $SO(N+1)$ with algebra
\begin{equation}
[L_{ab},L_{cd}]=i(\delta_{bc}L_{ad} + \delta_{ad}L_{bc} -\delta_{ac}L_{bd} -\delta_{bd}L_{ac}),
\end{equation}
where $L_{ab}=-L_{ba}$ (there is no distinction between upper and lower Euclidean indices in ${\bf R}^{N+1}$,
$X^a=X_a$).  This  can be extended to a representation of the Euclidean group,
$E_{N+1}$ acting on ${\bf R}^{N+1}$,
\begin{eqnarray}
[L_{ab},L_{cd}]&=&i(\delta_{bc}L_{ad} 
+ \delta_{ad}L_{bc} -\delta_{ac}L_{bd} -\delta_{bd}L_{ac})\\
\label{LL}
[L_{ab},X_c]&=&i(\delta_{bc}X_a-\delta_{ac}X_b) \\
\label{LX}
[X_a,X_b]&=& 0.
\label{XX}
\end{eqnarray}
An explicit realisation of this algebra in the continuum is
\begin{equation}
L_{ab}=i\left(X_a{\partial\over\partial X_b}-X_b{\partial\over\partial X_a}\right)
\end{equation}
acting on functions and $X_a$ being commutative multiplication
by the co-ordinates. 

The algebra will now be deformed, essentially using the inverse
of In\"on\"u-Wigner contraction. Let $X_a:=\mu L_{a,N+2}$,
with $\mu$ real,
and replace the commutator (\ref{XX}) above with
\begin{equation}
[X_a,X_b]=-i\mu^2L_{ab},
\label{XXL}
\end{equation}
leaving the other commutators unchanged.
The Euclidean group $E_{N+1}$ is thus deformed to $SO(N+2)$
\begin{equation}
[L_{AB},L_{CD}]=i(\delta_{BC}L_{AD} + \delta_{AD}L_{BC} -\delta_{AC}L_{BD} -\delta_{BD}L_{AC}),
\end{equation}
where $A,B,C,D=1,\ldots,N+2$ and $L_{a,N+2}=-L_{N+2,a}$.
In terms of the quadratic Casimirs\footnote{Relative to the standard 
conventions
for $SU(n)$ our normalisation here is such that 
$C_2^{SU(2)}={1\over 2}C_2^{3}$ and 
$C_2^{SU(4)}={1\over 2}C_2^{6}$.} 
$C_2^{N+1}=(1/2)L_{ab}L_{ab}$ and
$C_2^{N+2}=(1/2)L_{AB}L_{AB}$ we see that
\begin{equation}
X_aX_a={\mu^2}\left(C_2^{N+2}-C_2^{N+1}\right).
\label{XXR}
\end{equation}
While it may be tempting to think
of $X^a$ in (\ref{XXR}) as co-ordinates on a fuzzy sphere
we must be careful: a general irreducible representation of $SO(N+2)$
will decompose into a sum of different irreducible representations of \hbox{$SO(N+1)$},
with different values of $C_2^{N+1}$, so the right-hand side of (\ref{XXR})
will not be central.
However $X_aX^a$ {\it is} central in the fundamental spinor representation of 
$Spin(N+2)$.
To see this consider the even and odd cases separately:
\begin{itemize}
\item Even $N$: choose one chirality of spinor with $2^{N/2}$ components.
Under $Spin(N+2)\rightarrow Spin(N+1)$ this reduces uniquely to the single spinor 
representation
of $Spin(N+1)$, which has the same dimension.  In this case the right-hand side
of (\ref{XXR}) is a multiple of the identity.
\item Odd $N$: in this case the $2^{N/2}$ dimensional spinor representation
of $Spin(N+2)$ decomposes into two spinor representations of opposite chirality
under $SO(N+2)\rightarrow SO(N+1)$, both of dimension $2^{(N/2)-1}$.
But the second order Casimir $C_2^{N+1}$ has the same value
on the two chiralities, it does not distinguish between them, \cite{Fulton+Harris}.
So again the right-hand side of (\ref{XXR}) is a multiple of the identity.
\end{itemize}
The fundamental spinor representations of $Spin(k)$ have quadratic Casimir $C_2^{k}=k(k-1)/4$
for both even and odd $k$, see \cite{Fulton+Harris} for example.
For a fundamental spinor representation of $Spin(N+2)$ equation (\ref{XXR})
therefore gives
\begin{equation}
X_aX^a={\mu^2(N+1)\over 2}{\bf 1},
\end{equation}
where ${\bf 1}$ is the identity operator, and we might interpret $\sqrt{N+1\over 2}\mu$
as the radius of a fuzzy sphere. However, unlike $S^2_F$, the matrix
algebra generated by $X_a$ will not close in general even when $X_aX^a$ is
central.
Also one cannot get higher
dimensional representations of $S^N_F$ by taking tensor products.
This reflects the fact that there is no closed finite matrix approximation
to $S^N_F$ for $N\ne 2$.

Let us now investigate the geometry of the space generated by the adjoint
action of $Spin(N+2)$ on a fiducial direction $X_{N+1}$ (the ``north pole'' of
$S^N_F$).  Clearly $[L_{\alpha\beta}, X_{N+1}]=0$ for $\alpha,\beta=1,\ldots,N$,
so $Spin(N)$ leaves $X_{N+1}$ invariant.  Also $L_{N+1,N+2}$ commutes
with $X_{N+1}$, since they are just multiples of one another, so the  $SO(2)$ 
generated by $L_{N+1,N+2}$ also leaves $X_a$ invariant.  The upshot
of this is that the manifold that $D(g)\in Spin(N+2)$ generates with
the action
\begin{equation}
D^{-1}(g) X_{N+1} D(g)
\end{equation}
is the orthogonal Grassmannian $Q^N$,
which has dimension $2N$.
For $N\le 4$ these spaces have the following structures:
\begin{itemize}
 \item{} $Q^1\cong SO(3)/SO(2)\cong S^2$;
 \item{} $Q^2\cong SO(4)/[SO(2)\times SO(2)]\cong S^2\times S^2$
($S^2\times S^2$ was used in a consideration of fuzzy $S^3/{\bf Z}_2$
in \cite{Nair});
 \item{} $Q^3\cong SO(5)/[SO(3)\times SO(2)]\cong CP^3/{\bf Z}_2$
(this identification is described in \cite{S3});
 \item{} $Q^4\cong SO(6)/[SO(4)\times SO(2)]\cong SU(4)/[S(U(2)\times U(2))]$
(matrix approximations to this space were described in \cite{G2N}).
\end{itemize}

We propose to identify $Q^N$ 
with a \lq compactified' co-tangent bundle
for a non-commutative sphere.
This reduces to the usual $T^*S^N$ under In\"on\"u-Wigner contraction 
$\mu\rightarrow 0$:
this limit performs the dual function of rendering the 
$X$'s in (\ref{XXL})
commutative while at the same time 
de-compactifying \hbox{$SO(N+2)$} to the Euclidean group $E_{N+1}$.
 
\section{Fuzzy Complex Quadrics, $Q^N_F$}
Any unitary 
irreducible representation $T$ of a simple compact Lie group
is finite dimensional and the extension to
its enveloping algebra is a finite dimensional
matrix algebra.
This matrix algebra provides a fuzzy approximation to the algebra
of functions on the co-adjoint orbit of any Lie algebra element in $T$.

We shall refer to a finite matrix approximation to $Q^N$
as a fuzzy complex quadric and denote it by $Q^N_F$.
Being a co-adjoint orbit  $Q^N$ is a symplectic
manifold 
whose algebra of functions can be approximated by closed finite dimensional matrix
algebras. 
The harmonic expansion of a function on $Q^N$
requires all representations of $SO(N+2)$ which contain the trivial representation under $SO(N)\times SO(2)$.
A spinor representation of $SO(N+2)$ never contains a singlet of $SO(N)\times SO(2)$
for either even or odd $N$, so we restrict to vectorial representations.
The representation space of $SO(N+2)$ is then the space of rank-$n$
tensors, $T_{A_1\cdots A_n}$. Irreducible representations 
can be characterised by
their symmetries under interchange of their indices and
are traceless in any two indices. 
They can be represented by Young tableau with $n$ boxes
which reflect their permutation symmetries.

Any tensor that is anti-symmetric in three or more of its
indices must vanish when restricted to a representation of $SO(2)$
and hence cannot contribute to the harmonic expansion of
functions on $Q^N$.  Thus we can restrict our attention to
tensors that are anti-symmetric in pairs of indices only.
If $T_{A_1\cdots A_n}$ has $m$ pairs of anti-symmetric indices and 
$n-2m$ symmetric indices then the corresponding Young tableau is
\begin{equation}
\overbrace{\young(\ \Dots \ ,\ \Dots \  )}^m
\kern -2.2pt \raise 6.7pt\hbox{$\overbrace{\young(\ \Dots \ )}^{n-2m}$}\;.
\label{tableau}
\end{equation}
When $SO(N+2)$ is restricted to $SO(N)\times SO(2)$
the anti-symmetric pairs all contain singlets of $SO(N)\times SO(2)$
(when both indices in a pair are $SO(2)$ indices, for example).
Also if $n$ is even the $n-2m$ 
symmetric indices can contain singlets of $SO(N)\times SO(2)$,
but not when $n$ is odd. 
Thus the harmonic expansion of a function on $Q^N$ requires
all $SO(N+2)$ tensor representations of the form (\ref{tableau})
with $n$ even.
The dimension of these representation can be determined,
using the relevant formulae in \cite{Fulton+Harris} for example.  
With $n=2l$ they are, for $N\ge 3$,
\begin{eqnarray}
d_N(2l,m)&=&(2l+N-1)(2l+1-2m)(4l+N-2m)(N-2+2m)\nonumber\\
&&\qquad\qquad \times {(2l+N-2-m)!(N-3+m)!
\over N!(N-2)!(2l-m+1)!m!}
\end{eqnarray}
where $m\le l$.
\footnote{In the notation of \cite{Fulton+Harris} the representations
(\ref{tableau}) have highest weights 
$(r_1,\ldots,r_{[N/2]+1})=(n-m,m,0,\ldots,0)$, where $[N/2]$ is 
the integer part of $N/2$.}
If the harmonic expansion of a function on $Q^N$ is truncated
at $l_{max}=L$ the total number of degrees of freedom is
\begin{equation}
\sum_{l=0}^L\sum_{m=0}^ld_N(2l,m)=
\left[{(2L+N)(L+N-1)!\over L!N!}\right]^2=[d_N(L,0)]^2.
\label{matrixsize}
\end{equation}
The fact that this is a perfect square reflects the fact that
the degrees of freedom in a truncated harmonic expansion can
be arranged into a square matrix of size $d_N(L,0)\times d_N(L,0)$.
This can be represented in terms of Young tableau by
\begin{equation}
\overbrace{\young(\ \Dots \ )}^L
\times\overbrace{\young(\ \Dots \ )}^L
=1\oplus\young(\ ,\ )\oplus\young(\ \ )
\oplus \cdots \oplus\overbrace{\young(\ \Dots\ ,\ \Dots\ )}^L
\oplus\cdots\oplus \overbrace{\young(\ \Dots\ )}^{2L}
\label{MatDecomp}
\end{equation}
which relates the matrix structure, on the left-hand side, to
the harmonic expansion, on the right-hand side.
Matrix multiplication then gives a closed associative, but
non-commutative, algebra which reproduces the commutative
algebra of functions on $Q^N$ as 
$L\rightarrow\infty$.
At the level of functions the non-commutative product at finite $L$
can be realised as a $*$-product for the fuzzy complex quadric, $Q^N_F$.

\section{Functional Integrals on Fuzzy Spheres, $S^N_F$}

There is no closed finite matrix approximation
for the truncated algebra of functions on $S^N$ known, 
except for the special case
$N=2$ which was first described in \cite{Madore}. 
There does exist a matrix approximation to functions
on $S^N$, but in general matrix multiplication does not
correspond to the algebra of functions and the latter can only
be recovered by projecting back onto a function on $S^N$ after
matrix multiplication \cite{Ramgoolam}.  This results in
a non-associative algebra 
when $N\ne 2$ which, by a slight abuse of language, is nevertheless 
still referred to as a ``fuzzy'' sphere, $S^N_F$.

In the construction presented here a similar projection can be performed,
since the truncated harmonic expansion of a function on $S^N$ is buried
in $\QF$.  To see this note that functions on $S^N$ can be expanded
in symmetric tensor representations of $SO(N+1)$.  A truncation
at level $l_{max}$ requires using all symmetric tensors 
of $SO(N+1)$, $T_{a_1\cdots a_l}$,
with $0\le l\le l_{max}$. These are all contained in one
symmetric representation
\begin{equation}
\overbrace{\young(\ \Dots \ )}^{l_{max}}
\end{equation}
of $SO(N+2)$ under $SO(N+2)\rightarrow SO(N+1)$. 
Setting $l_{max}=2L$ we see that the last representation
on the right-hand side of (\ref{MatDecomp}) contains all the
$SO(N+1)$ representations necessary for the harmonic expansion
of a function on $S^N$ up to angular momentum $2L$.
We can thus obtain the fuzzy sphere $S^N_F$ by projecting
the irreducible representation $\overbrace{\young(\ \Dots\ )}^{2L}$ 
out from the matrix algebra of $\QF$ in (\ref{MatDecomp}).

This can be achieved in a functional integral over $\QF$ in the same
manner as in \cite{S3}.  Let $\Phi$ be a matrix in the algebra of $\QF$
for a given $L$.  
The $SO(N+2)$ invariant Laplacian on $\QF$ is
\begin{equation}
{\cal L}^2_{(N+2)}\Phi=-(1/2)[L_{AB},[L_{AB},\Phi]].
\end{equation}
Then the action for a scalar field on
$\QF$ can be written as
\begin{equation}
S[\Phi]={1\over d_N(L,0)}Tr\left\{\Phi^\dagger{\cal L}^2_{(N+2)}\Phi
+ V(\Phi)\right\}.
\label{action}
\end{equation}
with the scalar potential $V(\Phi^\dagger)=V(\Phi)$ assumed bounded below.
A functional integral then involves
\begin{equation}
Z=\int{\cal D}\Phi \hbox{e}^{-S[\Phi]}.
\end{equation}
We focus on the $S^N_F$ embedded in $\QF$ by penalising all the
$SO(N+2)$ representations in $Z$ except the last one on the right-hand 
of (\ref{MatDecomp}).  This can be achieved by modifying the
kinetic term in the action.
The second order Casimir of the representation (\ref{tableau}), with $n=2l$,
is \cite{Fulton+Harris}
\begin{equation}
C_2^{N+2}(2l,m)=(2l-m)(2l-m+N)+m(m+N-2).
\end{equation}
Observe that the completely symmetric tensors with $m=0$ have the
largest Casimir for any given $l$,
\begin{equation}
C_2^{N+2}(2l,0)=2l(2l+N).
\end{equation}
Hence the operator
\begin{equation}
-\left({1\over 2}\right)[L_{AB},[L_{AB},\cdot]-2L(2L+N)=
C_2^{N+2}-2L(2L+N)
\end{equation}
acting on $\Phi$ is negative for all modes in $\Phi$ 
except for the top mode,
with $l=L$ and $m=0$, on which it vanishes.
The $SO(N+1)$ invariant Laplacian on $S^N_F$ would be
\begin{equation}
{\cal L}^2_{(N+1)}\Phi=-(1/2)[L_{ab},[L_{ab},\Phi]].
\end{equation}
So the action
\begin{equation}
S_h[\Phi]=
{1\over d_N(L,0)}Tr\left\{\Phi^\dagger{\cal L}^2_{(N+1)}\Phi
+h\Phi^\dagger\left(-{\cal L}^2_{(N+2)}+2L(2L+N)\right)\Phi
+ V(\Phi)\right\},
\label{SNaction}
\end{equation}
with $h\gg1$,
will suppress all the unwanted modes in a functional
integral and leave the required modes for $S^N_F$ unaffected.  
In the limit $h\rightarrow \infty$ all modes, except
the ones relevant to $S^N_F$, will be suppressed and
correlation functions calculated with
\begin{equation}
Z=\int{\cal D}\Phi \hbox{e}^{-S_\infty[\Phi]}.
\end{equation}
will be those of the fuzzy sphere, truncated at level $2L$.

\section{Conclusions}
By constructing fuzzy approximations to complex quadrics,
$Q^N_F$, a prescription for defining functional integrals over
finite approximations to spheres has
been presented.  Although finite matrix approximations
to the algebra of functions on $N$-dimensional spheres 
are not known for $N\ne 2$, one can construct matrix 
approximations to the functions, but matrix multiplication then 
takes one out of the space of functions on the sphere.

The construction presented here relies in the fact that
the complex quadrics (which are orthogonal Grassmannians
$Q^N\cong SO(N+2)/[SO(N)\times SO(2)]$)
are co-adjoint orbits, and hence do have finite matrix approximations
to their algebra of functions, $\QF$. These spaces are related
to the co-tangent bundles $T^*S^N$ --- they are in a sense
compactified versions of the co-tangent bundles,
compactified at the expense of introducing a non-commutativity
on the sphere.   The algebra
of functions on $\QF$
contains the relevant degrees of freedom
for a truncated harmonic expansion of a function on $S^N$.
The functional integral for a field theory defined on $Q^N$
can be regularised in a manner that preserves the isometries
and avoids Fermion doubling \cite{FermionDoubling} by defining it over the fuzzy space
$\QF$.  By modifying the kinetic term and using the action (\ref{SNaction})
the degrees of freedom that are not relevant to the underlying
sphere can be prevented from contributing to the functional
integral and the result is a well defined, finite approximation
for the functional integral of a quantum field theory on $S^N$.
The correct algebra is ensured by restricting $\Phi$
to be matrices of size $d_N(L,0)$ given in (\ref{matrixsize})
and the continuum is recovered in the limit $L\rightarrow\infty$.
This construction is well suited to numerical evaluation.

\end{document}